\documentclass[aps,prl,twocolumn, superscriptaddress, floatfix]{revtex4}
\usepackage{graphicx,amsfonts,amsmath,color,epsfig,epstopdf, ulem}
\newcommand{\SU}[1]{\ensuremath{\mathrm{SU}( #1 )}}

\newcommand{\SpR}[1]{\ensuremath{\mathrm{Sp}( #1,\mathbb{R} )}}
\newcommand{\betb}{\begin{tabular}{p{4.0cm}p{9.0cm}}}
\newcommand{\entb}{\end{tabular}}

\begin{document}

\title{Collective Modes in Light Nuclei from First Principles}
\author{T. Dytrych}
\affiliation{Department of Physics and Astronomy, Louisiana State University, Baton Rouge, LA 70803, USA}
\author{K. D. Launey}
\affiliation{Department of Physics and Astronomy, Louisiana State University, Baton Rouge, LA 70803, USA}
\author{J. P. Draayer}
\affiliation{Department of Physics and Astronomy, Louisiana State University, Baton Rouge, LA 70803, USA}
\author{P. Maris}
\affiliation{Department of Physics and Astronomy, Iowa State University, Ames, IA 50011, USA}
\author{J. P. Vary}
\affiliation{Department of Physics and Astronomy, Iowa State University, Ames, IA 50011, USA}
\author{E. Saule}
\affiliation{Department of Biomedical Informatics, The Ohio State University, Columbus, OH 43210, USA}
\author{U. Catalyurek}
\affiliation{Department of Biomedical Informatics, The Ohio State University, Columbus, OH 43210, USA}
\affiliation{Department of Electrical and Computer Engineering, The Ohio State University, Columbus, OH 43210, USA}
\author{M. Sosonkina}
\affiliation{Department of Modeling, Simulation and Visualization Engineering, Old Dominion University, Norfolk, VA 23529, USA}
\author{D. Langr}
\affiliation{Department of Computer Systems, Czech Technical University in Prague, Prague, Czech Republic}
\author{M. A. Caprio}
\affiliation{Department of Physics, University of Notre Dame, Notre Dame, IN 46556, USA} 

\begin{abstract}
Results for \textit{ab initio} no-core shell model calculations in 
a symmetry-adapted \SU{3}-based coupling scheme 
demonstrate that collective modes in light nuclei emerge from first 
principles. The low-lying states of $^6$Li, $^8$Be, and $^6$He 
are shown to exhibit orderly patterns that favor spatial configurations 
with strong quadrupole deformation and complementary low intrinsic 
spin values, a picture that is consistent with the nuclear symplectic model. 
The results also suggest a pragmatic path forward to accommodate deformation-driven 
collective features in \textit{ab initio} analyses when they dominate the nuclear landscape.
\end{abstract}


\maketitle

\noindent
{\bf Introduction. --}
Major progress in the development of realistic inter-nucleon interactions
along with the utilization of massively parallel computing resources~\cite{HPC1, HPC2, HPC3} have placed
\textit{ab initio} approaches 
\cite{NCSM, BarrettNV13, MarisVN13, GFMC, CCM, NeffF04, NCSMstudies2, NCSMreactions2, BognerFMPSV08, RothLCBN11, EpelbaumKLM11}
at the frontier of nuclear structure explorations. The ultimate goal of
\textit{ab initio} studies is to
establish a link between underlying principles of quantum chromodynamics
(quark/gluon considerations) and observed properties of atomic nuclei, including their
structure and related reactions. The predictive potential that \textit{ab initio}
models hold 
\cite{MarisSV10, Goldberg} 
makes them suitable for targeting short-lived nuclei that are
inaccessible by experiment but essential to modeling, for example,  
of the dynamics of X-ray bursts and the path of nucleosynthesis 
(see, e.g., \cite{DavidsCJM11, Laird13}).

In this letter, we report on \textit{ab initio} symmetry-adapted no-core shell
model (SA-NCSM) results for the $^{6}$Li (odd-odd), $^{8}$Be (even-even), and
$^{6}$He (halo) nuclei, using two realistic nucleon-nucleon ($NN$)
interactions, the JISP16~\cite{ShirokovMZVW04} and chiral N$^3$LO~\cite{EntemM03}
potentials.  The SA-NCSM framework exposes a remarkably simple physical feature that is
typically masked in other \textit{ab initio} approaches; the
emergence, without {\it a priori} constraints, of  simple orderly patterns that
favor spatial configurations with strong quadrupole deformation and  low
intrinsic spin values. This feature, once exposed and understood, can 
be used to guide a truncation and augmentation of model spaces to ensure that 
important properties 
of atomic nuclei, like enhanced $B(E2)$ strengths,
nucleon cluster substructures, and others important in reactions, are
appropriately accommodated in future \textit{ab initio} studies. 

The SA-NCSM joins a no-core shell model (NCSM) theory~\cite{NCSM} with
a multi-shell, \SU{3}-based coupling scheme \cite{Elliott, SANCSM}.
Specifically, nuclear wavefunctions are
represented as a superposition of many-particle configurations carrying a
particular intrinsic quadrupole deformation linked to the irreducible
representation (irrep) labels $(\lambda\,\mu)$ of \SU{3} \cite{RosensteelR77,
LeschberD87, SU3toShape}, and specific intrinsic spins $(S_{p}S_{n}S)$ for
protons, neutrons, and total spin, respectively (proton-neutron
formalism). The fact that  \SU{3} plays a key role, e.g., in the
microscopic description of the experimentally observed collectivity of
$ds$-shell nuclei~\cite{Draayer25MgAl, Reactions, DraayerReactions,
DraayerWR84, Sp3R24Mg}, and for heavy deformed systems~\cite{DraayerW83},
tracks from the seminal work of Elliott \cite{Elliott} and is 
reinforced by the fact that it is the underpinning symmetry of the microscopic
symplectic model~\cite{Sp3R, RosensteelR80}, which provides a comprehensive
theoretical foundation for understanding the dominant symmetries of nuclear
collective motion \cite{RoweRPP, DraayerWR84}. 

\begin{figure*}
\includegraphics[width=0.49\textwidth]{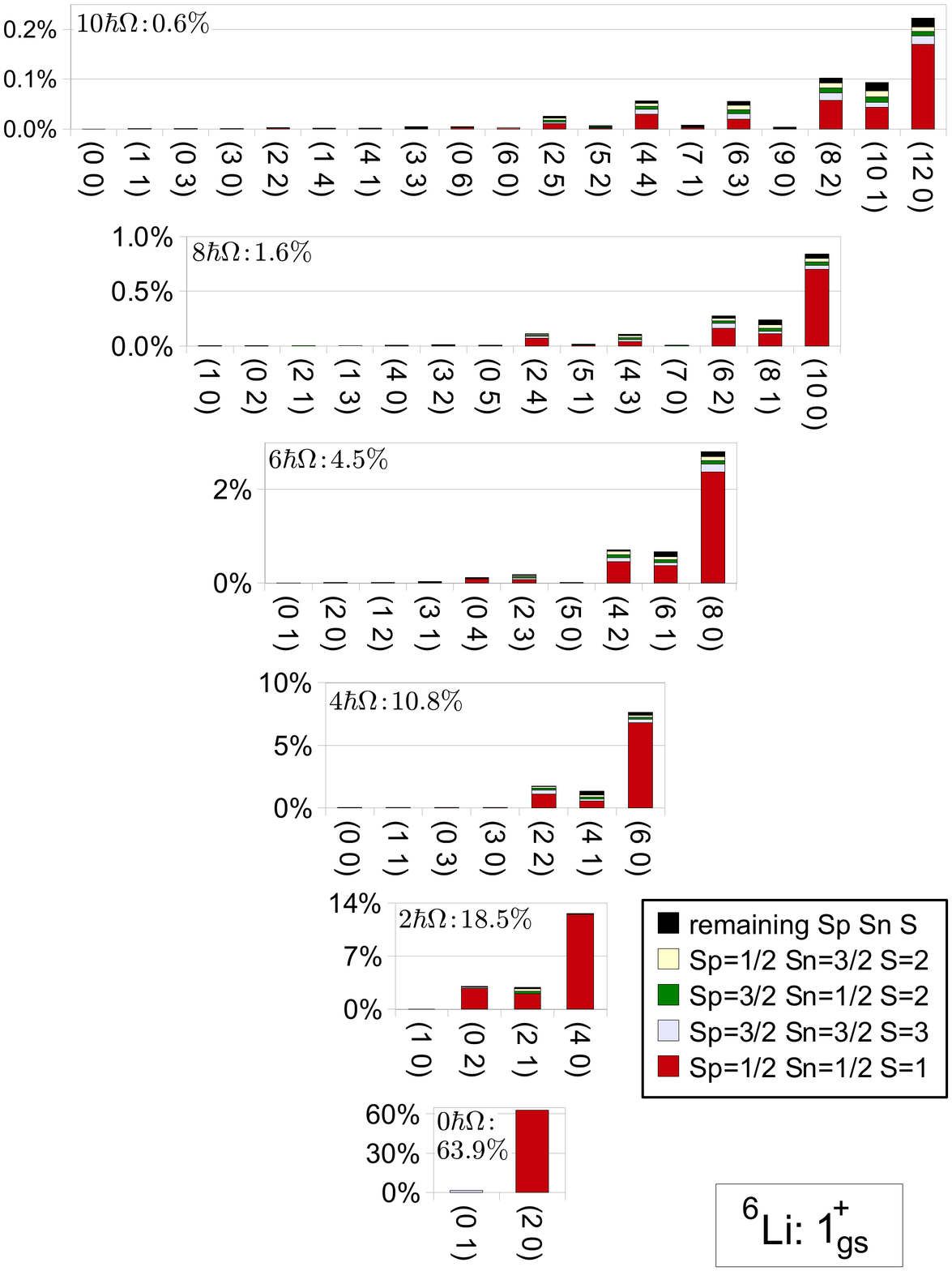} 
\includegraphics[width=0.49\textwidth]{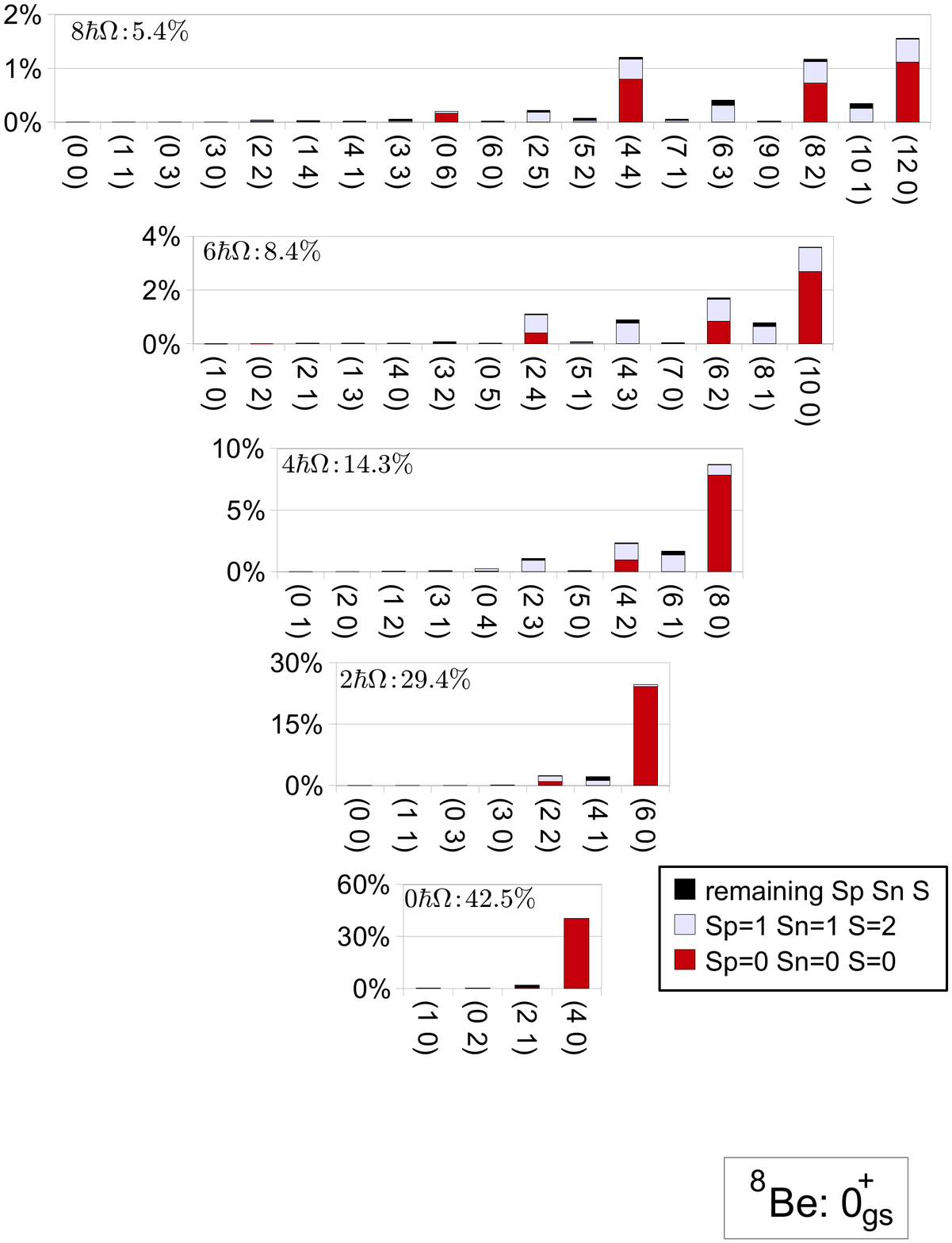} 
\vspace{-0.30cm}
\caption
{
	Probability distributions  across ($S_{p} S_{n} S$)  and $(\lambda\,\mu)$
	values (horizontal axis) for the calculated $1^{+}_{gs}$  of
	$^{6}$Li obtained for $N_{\max}=10$ and $\hbar\Omega=20$ MeV with the
	JISP16 interaction (left) and the $0^{+}_{gs}$  of $^{8}$Be
	obtained for $N_{\max}=8$ and $\hbar\Omega=25$ MeV with the chiral N$^3$LO 
	interaction (right). The total probability for each $N\hbar\Omega$ subspace
	is given in the upper left-hand corner of each histogram.  The concentration of
	strengths to the far right  demonstrates the
	dominance of collectivity.
}
\label{fig:gsStructure}
\vspace{-0.5cm}
\end{figure*}

The outcome further suggests a symmetry-guided basis selection that yields
results that are nearly indistinguishable from the complete basis counterparts.
This is illustrated for $^{6}$Li and $^{6}$He for a range of harmonic
oscillator (HO) energies $\hbar\Omega$, and $N_{\max}$=12 model spaces, where
$N_{\max}$ is the maximum number of HO quanta included in the basis states
above the Pauli allowed minimum for a given nucleus. An overarching long-term
objective is to extend the reach of the standard NCSM scheme by exploiting
symmetry-guided principles that enable one to include configurations beyond the
$N_{\max}$ cutoff, while capturing the essence of long-range
correlations that often dominate the nuclear landscape.

\noindent
{\bf  {\it Ab initio} realization of collective modes. --}
The expansion of eigenstates in the physically relevant SU(3) basis
unveils salient features that emerge from the complex dynamics of  these
strongly interacting many-particle systems.  To explore the nature of the most
important correlations, we analyze the probability distribution across
$(S_{p} S_{n} S)$ and $(\lambda\,\mu)$ configurations of the four
lowest-lying isospin-zero ($T=0$) states of $^6$Li ($1^+_{\rm gs}$, $3^+_1$,
$2^+_1$, and $1^+_2$), along with the ground-state rotational bands of $^{8}$Be
and $^{6}$He.  Results for the ground-state of $^{6}$Li and $^{8}$Be, obtained
with the JISP16 and chiral N$^3$LO interactions, respectively, are shown in
Figure~\ref{fig:gsStructure}.  This figure illustrates a feature common to all
the low-energy solutions considered; namely, a highly structured and regular
mix of intrinsic spins and \SU{3} spatial quantum numbers that has heretofore
gone unrecognized in other \textit{ab initio} studies, and which 
does not seem to depend on the particular choice of realistic $NN$ potential.  

First, consider the spin content.  
 The calculated eigenstates project at a 99\% level onto a 
comparatively small subset of intrinsic spin combinations.  For
instance, the lowest-lying eigenstates in $^{6}$Li are almost entirely realized
in terms of configurations characterized by the following intrinsic spin
$(S_{p}S_{n}S)$ triplets: $(\frac{3}{2} \frac{3}{2} 3),\,\, (\frac{1}{2}
\frac{3}{2} 2),\,\,(\frac{3}{2}\frac{1}{2} 2)$, and $(\frac{1}{2} \frac{1}{2}
1)$, with the last one carrying over 90\% of each eigenstate.  Similarly, the
ground-state bands of $^{8}$Be and $^{6}$He are found to be dominated by 
configurations carrying total intrinsic spin of the protons and neutrons equal
to zero and one, with the largest contributions due to $(S_{p}S_{n}S)=(000)$ and $(112)$
configurations.

Second, consider  the spatial degrees of freedom.  The mixing of
$(\lambda\,\mu)$ quantum numbers exhibits a remarkably simple pattern.  One of its
key features is the preponderance of a single $0\hbar\Omega$ \SU{3} irrep. 
This so-called leading irrep, is characterized by the largest value of the
intrinsic quadrupole deformation~\cite{SU3toShape}. For instance, the low-lying states of
$^{6}$Li project at a 40\%-70\% level onto the prolate $0\hbar\Omega$ \SU{3}
irrep $(2\,0)$, as illustrated in Fig.~\ref{fig:gsStructure}.  For the ground-state band of $^{8}$Be and $^{6}$He, qualitatively
similar dominance of the leading $0\hbar\Omega$ \SU{3} irreps is observed. The dominance of the most
deformed $0\hbar\Omega$ configuration  indicates that the quadrupole-quadrupole interaction
of the Elliott \SU{3} model~\cite{Elliott} is
realized naturally within an {\textit{ab initio}} framework.

The analysis also reveals that the dominant \SU{3} basis states at each
$N\hbar\Omega$ subspace ($N=0, 2, 4,\dots$) are typically those with
$(\lambda\, \mu)$ quantum numbers given by
\begin{equation}
\lambda + 2\mu = \lambda_{0} + 2\mu_{0} + N	
\label{eq:Sp2RSelection}
\end{equation}
where $\lambda_{0}$ and $\mu_{0}$ denote labels of the leading \SU{3} irrep 
in the $0\hbar\Omega$ ($N=0$) subspace. 
We conjecture that this regular pattern of \SU{3} quantum numbers
reflects the presence of an underlying symplectic \SpR{3} symmetry of
microscopic nuclear collective motion \cite{Sp3R} that governs the low-energy
structure of both even-even and odd-odd $p$-shell nuclei. This can be seen from
the fact that $(\lambda\, \mu)$ configurations that satisfy
condition~(\ref{eq:Sp2RSelection}) can be determined from the leading \SU{3}
irrep $(\lambda_{0}\, \mu_{0})$ through a successive application of a specific
subset of the \SpR{3} symplectic $2\hbar\Omega$ raising operators.  This subset
is composed of the three operators, $\hat{A}_{zz}, \hat{A}_{zx}$, and
$\hat{A}_{xx}$, that distribute two oscillator quanta in $z$ and $x$
directions, but none in $y$ direction, thereby inducing \SU{3} configurations
with ever-increasing intrinsic quadrupole deformation.  These three operators
are the generators of the $\SpR{2}\subset \SpR{3}$
subgroup~\cite{PetersonHecht}, and give rise to deformed shapes
that are energetically favored by an attractive quadrupole-quadrupole
interaction~\cite{RoweRPP}. This is consistent with our earlier
findings of a clear symplectic \SpR{3} structure with the same pattern
(\ref{eq:Sp2RSelection}) in \textit{ab initio} eigensolutions for $^{12}$C and
$^{16}$O~\cite{Dytrych}.

Furthermore, the $N\hbar\Omega$
configurations with $(\lambda_{0}\!+\!N\,\,\mu_{0})$, the so-called stretched
states, carry a noticeably higher probability than the others. For instance,
the $(2\!+\!N\,\,0)$ stretched states contribute at the 85\% level to the
ground-state of $^{6}$Li, as can be readily seen in Fig.~\ref{fig:gsStructure}.
The sequence of the stretched states is formed by consecutive applications of
the $\hat{A}_{zz}$ operator, the generator of
$\SpR{1}\subset\SpR{2}\subset\SpR{3}$ subgroup, over the leading \SU{3} irrep.
This translates into distributing $N$ oscillator quanta along the direction of
the $z$-axis only and hence rendering the largest possible deformation.

\noindent
{\bf Symmetry-guided framework.} -- 
The observed patterns of intrinsic spin and deformation mixing supports the
symmetry-guided basis selection philosophy referenced above.  Specifically, 
one can take advantage of dominant symmetries to relax and refine the 
definition of the SA-NCSM model space, which for the NCSM is fixed by 
simply specifying the $N_{\max}$ cutoff.  
In particular,  SA-NCSM model spaces can be characterized by a pair of numbers, 
$\langle N^{\bot}_{\max} \rangle N^{\top}_{\max}$, which implies inclusion of the complete  
space up through $N^{\bot}_{\max}$, and a subset of the complete set of $(\lambda\,\mu)$ and 
$(S_p S_n S)$ irreps between $N^{\bot}_{\max}$ and $N^{\top}_{\max}$.
Though not a primary focus of this paper, an ultimate goal is to be able to carry out 
SA-NCSM investigations in deformed nuclei with $N^{\top}_{\max}$ values that go 
beyond the highest $N_{\max}$  for which complete NCSM results can be provided.

The SA-NCSM concept
focuses on retaining the most important configurations that support the
strong many-nucleon correlations of a nuclear system using underlying
$\SpR{1}\subset\SpR{2}\subset\SpR{3}$ symmetry considerations.  It is important to note that for model spaces truncated according
to $(\lambda\,\mu)$ and $(S_{p}S_{n}S)$ irreps, the
spurious center-of-mass motion can be factored out exactly~\cite{Verhaar}, 
which represents an important advantage of this scheme. 

The efficacy of the symmetry-guided concept is illustrated for SA-NCSM results
obtained in model spaces which are expanded beyond a
complete $N^{\bot}_{\max}$ space with irreps that span a relatively few dominant intrinsic
spin components and carry quadrupole deformation specified by
~(\ref{eq:Sp2RSelection}).  Specifically, we vary $N^{\bot}_{\max}$ from 2 to 10 with 
only the subspaces determined by (1) included beyond $N^{\bot}_{\max}$.
This allows us to study convergence of spectroscopic properties towards results obtained in the
complete $N_{\max}=12$ space and hence, probes the efficacy of the SA-NCSM
symmetry-guided model space selection concept.
In the present study, a Coulomb plus JISP16 $NN$ interaction for $\hbar\Omega$
values ranging from $17.5$ up to $25$ MeV is used, along with the
Gloeckner-Lawson prescription~\cite{Lawson} for elimination of spurious
center-of-mass excitations.  SA-NCSM eigenstates are used to determine
spectroscopic properties of low-lying $T=0$ states of $^{6}$Li and the
ground-state band of $^{6}$He for $\langle N^{\bot}_{\max}\rangle$12 model
spaces. 

\begin{figure}
\vspace{-0.4cm}
\includegraphics[width=0.5\textwidth]{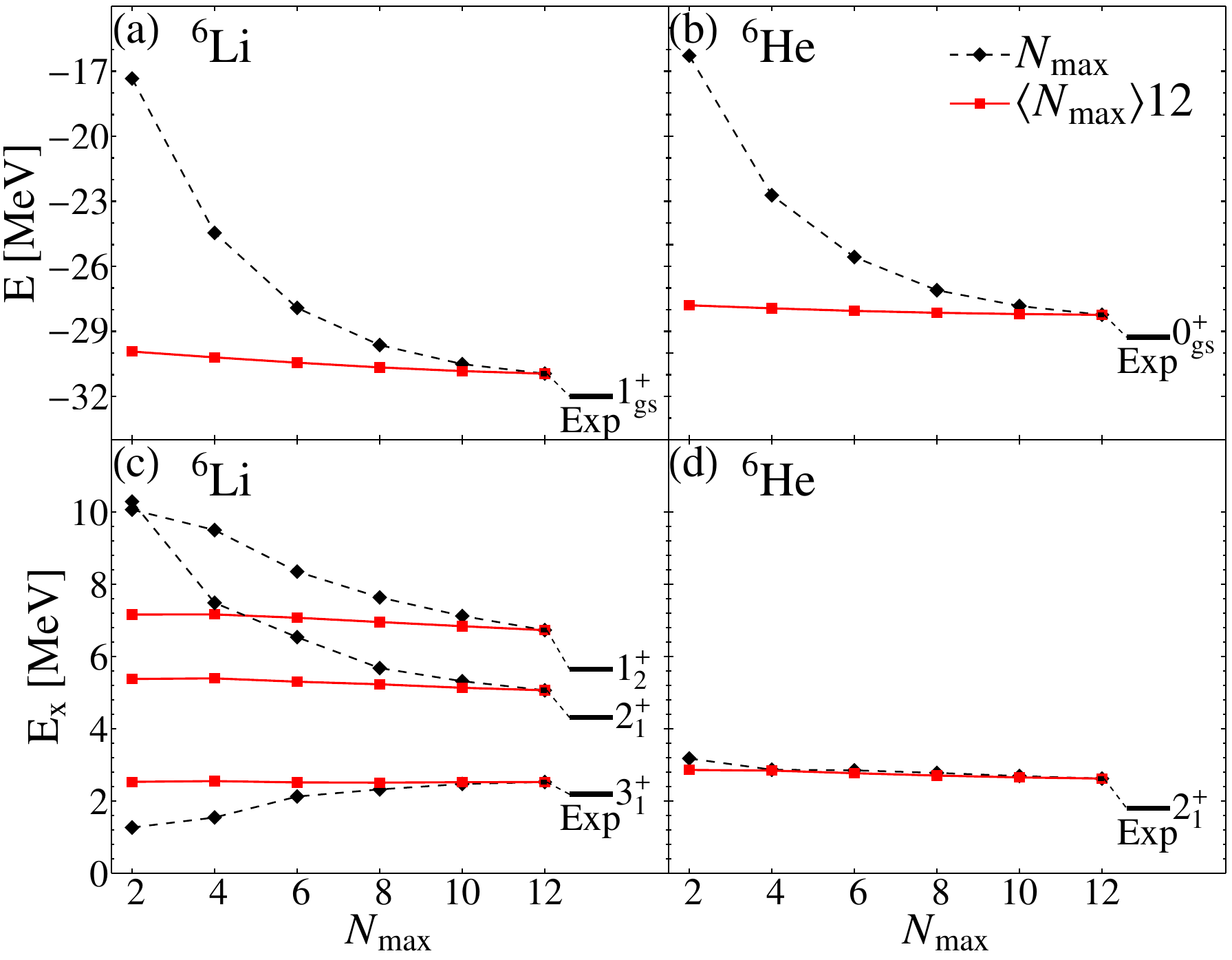}
\vspace{-0.25cm}
\caption{
The ground-state binding energies of $^{6}$Li (a) and $^{6}$He (b),
excitation energies of $T=0$ states of $^{6}$Li (c), $2^{+}_{1}$ excited state
of $^{6}$He (d), shown for the complete $N_{\max}$ (dashed black curves) and 
truncated $\langle N^{\bot}_{\max}=N_{\max}\rangle 12$ 
(solid red lines) model spaces.  Results shown are for JISP16 and $\hbar\Omega=20$ MeV.  
Note the relatively large changes when the complete space is increased from $N_{\max}=2$ to $N_{\max}=12$
as compared to nearly constant $\langle N_{\max} \rangle12$ SA-NCSM outcomes.
}
\label{6LiT0Spectrum}
\end{figure}

\begin{figure}[b!]
\vspace{-0.20cm}
\includegraphics[width=0.48\textwidth]{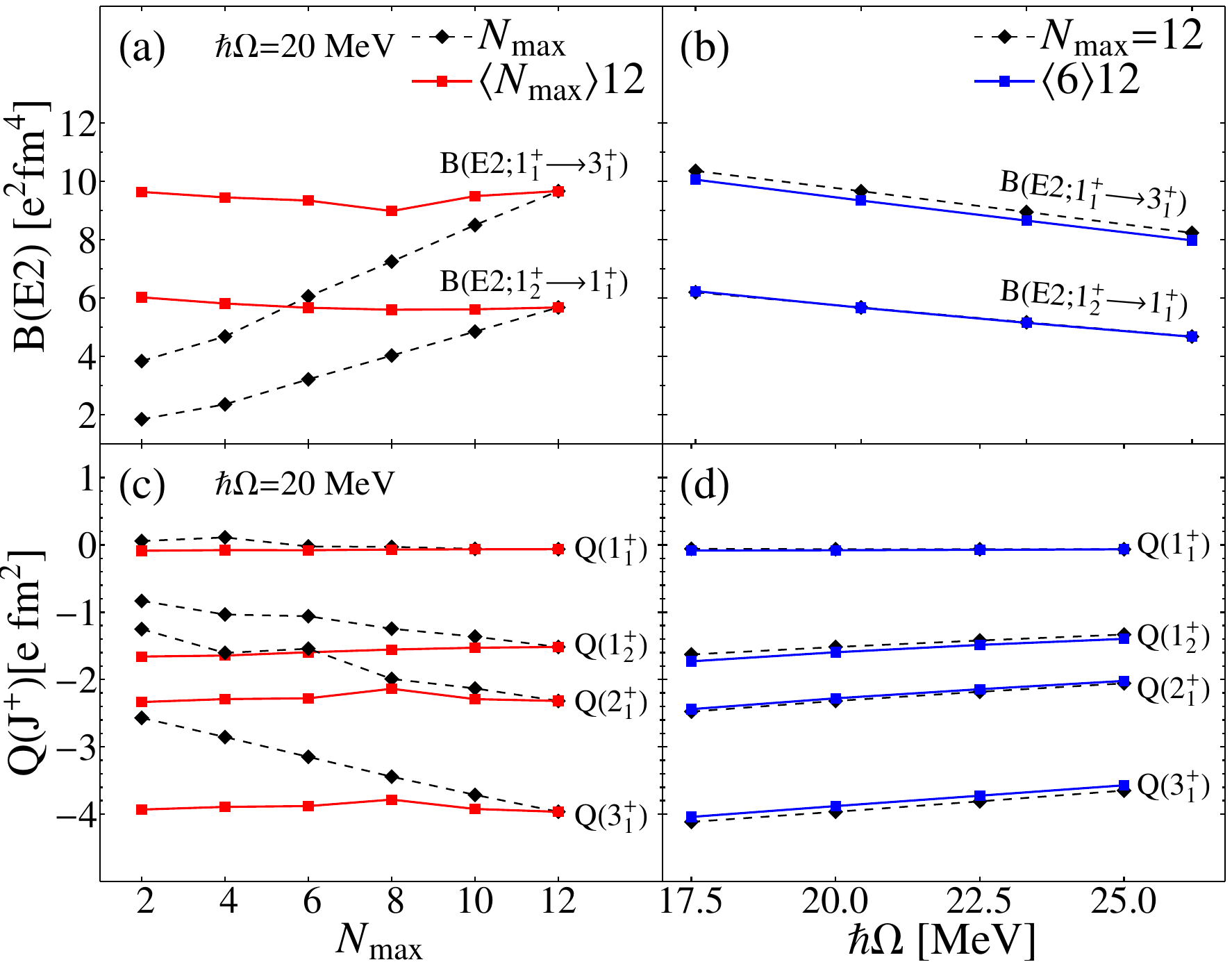}
\vspace{-0.20cm}
\caption{
Electric quadrupole transition probabilities and quadrupole moments for $T=0$
states of $^{6}$Li calculated using the JISP16 interaction without using effective charges 
are shown for the
complete $N_{\max}$ (dashed black lines) and truncated $\langle N^{\bot}_{\max}=N_{\max}\rangle
12$ (solid red lines) model spaces [(a) and (c)], and as a function of
$\hbar\Omega$  for the complete $N_{\max}=12$ space and $\langle6\rangle 12$
truncated space (solid blue lines) [(b) and (d)].
Experimentally, $B(E2;1^{+}_{1}\rightarrow
3^{+}_{1})=25.6(20)\,e^2\mathrm{fm}^4$ \cite{Tilley2002}.
}
\label{fig:BE2Q2}
\end{figure}

The results indicate that the observables obtained in the $\langle
N^{\bot}_{\max}\rangle$12 symmetry-guided truncated spaces are excellent
approximations to the corresponding $N_{\max}=12$ complete-space counterparts.
Furthermore, the level of agreement achieved is only marginally dependent
on $N_{\max}^{\bot}$. In particular, the ground-state binding energies
obtained in a $\langle 2\rangle 12$ model space represent
approximately 97\% of the complete-space $N_{\max}=12$ binding energy in
the case of $^{6}$Li and reach over 98\% for $^{6}$He
[Fig.~\ref{6LiT0Spectrum} (a) and (b)].  The excitation energies
differ only by 5 keV to a few hundred keV from the corresponding
complete-space $N_{\max}=12$ results [see Fig.~\ref{6LiT0Spectrum} (c) and
(d)], and the agreement with known experimental data is reasonable over a broad
range of $\hbar\Omega$ values.

The number of basis states used, e.g., for each $^6$Li state, is
only about 10-12\% for $\langle 2\rangle 12$, $\langle 4\rangle 12$, $\langle
6\rangle 12$, 14\% for $\langle8\rangle 12$, and 30\% for $\langle 10\rangle
12$ as compared to the number for the complete $N_{\max}=12$ model
space, which is $3.95\times 10^{6}$ ($J=1$), $5.88\times 10^{6}$ ($J=2$), and
$6.97\times 10^{6}$ ($J=3$). The runtime of the SA-NCSM code exhibits a quadratic
dependence on the number of  $(\lambda\,\mu)$ and  $(S_p S_n S)$ irreps for a nucleus --
there are $1.74\times10^{6}$ irreps for the complete $N_{\max}$ = 12 model
space of $^{6}$Li, while only 8.2\%, 8.3\%, 8.9\%, 12.7\%, and 30.6\% of these are
retained for $N^{\bot}_{\max} = 2, 4, 6, 8,$ and $10$, respectively. The net
result is that calculations in the $10 \ge N_{\max}^{\bot} \ge 2$ range
require one to two orders of magnitude less
time than SA-NCSM calculations for  the complete $N_{\max}=12$
space.


\begin{table}
\vspace{-0.5cm}
\caption{
Magnetic dipole moments $\mu$ [$\mu_{N}$] and point-particle rms matter
radii $r_m$ [fm] of $T=0$ states of $^{6}$Li calculated in the complete $N_{\max}=12$ space and
the $\langle$6$\rangle$12 subspace for JISP16 and $\hbar\Omega=20$ MeV.  The experimental value for
the $1^+$ ground-state  is  known to be $\mu=+0.822$ $\mu_{N}$ \cite{Tilley2002}.
\label{tab:muRMS}}
\begin{ruledtabular}
\begin{tabular}{llcrrrr}
& & & $1^{+}_{1}$ & $3^{+}_{1}$ & $2_{1}^{+}$ & $1^{+}_{2}$ \\
$\mu$ & $N_{\max}=12$&    & 0.838 & 1.866 & 0.970 & 0.338  \\
& $\langle$6$\rangle$12 & & 0.839 & 1.866 & 1.014 & 0.338  \\
$r_m$ & $N_{\max}=12$ &   & 2.119 & 2.063 & 2.204 & 2.313  \\
& $\langle$6$\rangle$12 & & 2.106 & 2.044 & 2.180 & 2.290  \\
\end{tabular}
\end{ruledtabular}
\end{table}

\begin{table}
\vspace{-0.5cm}
\caption{
Selected observables for the two lowest-lying states of $^{6}$He obtained in
the complete $N_{\max}=12$ space and $\langle$8$\rangle$12 model subspace for JISP16 and $\hbar\Omega=20$ MeV. 
\label{tab:He6obs}}
\begin{ruledtabular}
\begin{tabular}{lcc}
& $N_{\max}=12$ & $\langle$8$\rangle$12 \\
$B(E2;2^{+}_{1}\rightarrow 0^{+}_{1})$ [$e^2$fm$^4$] &  0.181 &  0.184 \\
$Q(2^{+}_{1})$ [$e$fm$^2$] & -0.690 & -0.711 \\
$\mu(2^{+}_{1})$ [$\mu_{N}$]   & -0.873 & -0.817 \\
$r_m$ $(2^{+}_{1})$ [fm]          &  2.153 &  2.141 \\
$r_m$ $(0^{+}_{1})$ [fm]          &  2.113 &  2.110 \\
\end{tabular}
\end{ruledtabular}
\end{table}

As illustrated in Table~\ref{tab:muRMS}, the magnetic dipole moments
obtained in the $\langle 6\rangle 12$ model space for $^{6}$Li agree to
within 0.3\% for odd-$J$ values, and 5\% for $\mu(2^{+}_{1})$.  Qualitatively
similar agreement is achieved for $\mu(2^{+}_{1})$ of $^{6}$He, as shown in
Table~\ref{tab:He6obs}.  The results suggest that it may suffice to include all
low-lying $\hbar\Omega$ states up to a fixed limit, e.g. $N^{\bot}_{\max}=6$
for $^{6}$Li and $N^{\bot}_{\max}=8$ for $^{6}$He, to account for the most
important correlations that contribute to the magnetic dipole moment. 

To explore how close one comes to reproducing the important long-range
correlations, we compared observables that are sensitive to the tails of the
wavefunctions; specifically, the point-particle root-mean-square (rms) matter
radii, the electric quadrupole moments and the reduced electromagnetic $B(E2)$
transition strengths that could hint at rotational features \cite{CaprioPV13}.
As Table~\ref{tab:He6obs} shows, the complete-space $N_{\max}=12$ results for
these observables are remarkably well reproduced by the SA-NCSM for $^{6}$He in
the restricted $\langle$8$\rangle$12 space.  In addition, the results for the
rms matter radii of $^{6}$Li, listed in Table~\ref{tab:muRMS}, agree to within
1\% for the $\langle 6\rangle 12$ model space.

Notably, the $\langle 2 \rangle 12$ eigensolutions for $^{6}$Li yield results
for $B(E2)$ strengths and quadrupole moments that track closely with their
complete $N_{\max}=12$ space counterparts (see Fig.~\ref{fig:BE2Q2}).  It
is known that further expansion of the model space beyond
$N_{\max}=12$ is needed to reach convergence~\cite{Cockrell:2012vd,
PMarisConvergence}. However, the close correlation between the
$N_{\max}=12$ and $\langle$2$\rangle$12 results is 
strongly suggestive that this convergence can be obtained through the
leading \SU{3} irreps in a symmetry-adapted space.
In addition, the results [Fig.~\ref{fig:BE2Q2} (c)] reproduce 
the ground-state quadrupole moment~\cite{Qfootnote} that is
measured to be $Q (1^+) = -0.0818(17)\,e\mathrm{fm}^2$ \cite{Tilley2002}.

The differences between truncated-space and complete-space results are found to
be essentially $\hbar\Omega$ insensitive  and appear
sufficiently small as to be nearly inconsequential relative to the 
dependences on $\hbar\Omega$ and on $N_{\max}$ [see Fig.~\ref{fig:BE2Q2} (b) and (d)].
Since the NN interaction dominates contributions from three-nucleon forces
(3NFs) in light nuclei, except for selected cases~\cite{BarrettNV13,
NCSMstudies2, MarisVN13}, we expect our results to be robust and carry forward
to planned applications that will include 3NFs.

To summarize, the results reported in this paper demonstrate that observed
collective phenomena in light nuclei emerge naturally from first-principle considerations. This  is illustrated through detailed calculations
in a SA-NCSM framework for $^6$Li, $^{6}$He, and $^{8}$Be nuclei using the 
JISP16 and chiral N$^{3}$LO $NN$ realistic interactions.  The results underscore the
strong dominance of configurations with large deformation and low spins. 
The results also suggest a path forward to include higher-lying
correlations that are essential to collective features such as enhanced B(E2)
transition strengths.  The results further anticipate the 
significance of $LS$-coupling and \SU{3} as well as an underlying symplectic
symmetry for an extension of {\it ab initio} methods to the heavier,
strongly deformed nuclei of the lower $ds$ shell, and, perhaps, even reaching
beyond. 

We thank David Rowe and Andrey Shirokov for useful discussions. This work was
supported in part by the US NSF [OCI-0904874, OCI-0904809, PHY-0904782], the US
Department of Energy [DESC0008485, DE-FG02-95ER-40934, DE-FG02-87ER40371], the
National Energy Research Scientific Computing Center [supported by DOE’s Office
of Science under Contract No.  DE-AC02-05CH1123], the Southeastern Universities
Research Association, and by the Research Corporation for Science Advancement
under a Cottrell Scholar Award.  This work also benefitted from computing
resources provided by the Louisiana Optical Network Initiative and Louisiana
State University's Center for Computation \& Technology.  T. D. and D. L.
acknowledge support from Michal Pajr and CQK Holding.


\begin{thebibliography}{1}
\bibitem{Qfootnote}
The near vanishing of the quadrupole moment, an L=2 object, in the ground-state
of $^6$Li, $Q(1^{+}_{1})$, can be attributed to a very strong dominance ($\sim$87\%)
of L=0 configurations in the ground state.
\end{thebibliography}
\end{document}